\documentstyle[prl,aps]{revtex}
\newcommand{\be}{\begin{equation}}
\newcommand{\ee}{\end{equation}}
\newcommand{\ba}{\begin{eqnarray}}
\newcommand{\ea}{\end{eqnarray}}

\begin{document}
\draft
\title{Induced Parity Breaking Term at Finite Temperature}
\author{C.D.\ Fosco$^a$\thanks{CONICET}\,,
G.L.\ Rossini$^{b,c}$\thanks{CONICET. On leave from La Plata University, 
Argentina}\,
and\,
F.A.\ Schaposnik$^c$\thanks{Investigador CICBA, Argentina}
\\
{\normalsize\it
$^a$Centro At\'omico Bariloche,
8400 Bariloche, Argentina}\\
{\normalsize\it
$^b$ Center for Theoretical Physics, Laboratory for 
Nuclear Science and Department of Physics}\\
{\normalsize\it Massachusetts Institute of Technology,
 Cambridge, Massachusetts 02139, USA
}\\
{\normalsize\it
$^c$Departamento de F\'\i sica, Universidad Nacional de La Plata}\\
{\normalsize\it
C.C. 67, 1900 La Plata, Argentina}}

\maketitle
\begin{abstract}
We compute the exact induced parity-breaking part of the effective action
for $2+1$ massive fermions in $QED_3$  at finite temperature
by calculating the fermion determinant in a particular background.
The result confirms that gauge invariance of the effective
action is respected even when large gauge transformations are considered.
\end{abstract}


\bigskip


Because of its relevance both in Field Theory and
Condensed Matter physics, much effort has been devoted in the
last decade to the study of
three dimensional gauge theories  coupled
to matter. An important ingredient in this
theories  is  the parity anomaly \cite{djt}
which induces, through fluctuation of massive Fermi fields,
a Chern-Simons term in the effective action
of the gauge field \cite{det}.

As originally stressed in \cite{djt},
a fundamental property of the Chern-Simons (CS)  action is the existence
of a
quantization law: due to the non-invariance of
the CS term under gauge transformations of "non-zero" winding number,
the coefficient which appears in front of the
(non-Abelian) Chern-Simons three form $S_{CS}$
should be quantized so that $\exp(iS_{CS})$ is singled valued. Even in
the Abelian case ``large'' gauge transformations come into play whenever
the theory is formulated in an appropriately compactified manifold
\cite{polyNP}-\cite{hosotani} and in that case the quantization law also
holds. Putting together all these facts one can
state that in three dimensional gauge theories with Fermi
fields, calculations of the effective
action for the gauge field using  gauge-invariant
regularizations lead
to a parity anomaly which manifests through
the occurrence of a CS term with a  quantized coefficient which
depends on the number of fermion species.

A natural question raised when  the analysis of three dimensional
gauge theories was extended to
the case of finite temperature was whether quantization of
the CS coefficient induced by fermion fluctuations
survives the
effects of temperature or it is smoothly renormalized.
The question
was originally discussed in \cite{pis} where it was argued that
 the coefficient
of the CS term in the effective action for the gauge field should
remain unchanged at finite temperature. Contrasting with this analysis,
perturbative calculations for both relativistic and non-relativistic
theories, abelian and non-abelian, have yielded induced actions with
CS coefficients which are smooth functions of the temperature
\cite{NS}-\cite{I}, this seeming to signal a kind
of gauge anomaly at finite
temperature.

The problem  of renormalization
of the CS coefficient induced by fermions at $T \ne 0$
was revisited in refs.\cite{bfs}-\cite{cfrs} where it was
concluded that, on gauge invariance grounds and 
in perturbation theory,
the effective action for the gauge field
cannot contain a smoothly renormalized CS coefficient  at
non-zero temperature. More recently, the exact result for
the effective action of  a $0+1$ analogue of the
CS system \cite{dll} as well as a zeta-function analysis of the fermion
determinant at $T \ne 0$ in the $2+1$ model \cite{dgs} have explicitly
shown that although the perturbative expansion leads to a 
smooth temperature
dependent and hence non-quantized CS coefficient, the complete effective
action can be made gauge invariant, the induced CS term's non-invariance
revealed by perturbation theory
being compensated by non-local contributions to the effective action.

Originally \cite{det} the  parity anomaly for fermions at $T = 0$
was analyzed
by considering a particular gauge field background
configuration which allowed
the closed computation of the anomalous part of the fermion current.
In the same vein, we compute in this work the induced parity-breaking
part of the effective action for
three dimensional massive fermions in $QED_3$ at finite temperature
by considering a particular gauge field configuration
which allows the obtention of a closed exact result
for the fermion determinant. Our
result confirms that gauge invariance,
even under large gauge transformations, is respected and at the
same time reproduces in the appropriate limits the perturbative
and zero temperature results.

We define the total effective action $\Gamma (A)$, as usual, by the
formula
\be
e^{- \Gamma (A,M)} \;=\;  \int {\cal D} \psi \,
{\cal D} {\bar \psi} \; \exp \left[ - \int_0^\beta d \tau \int d^2 x \; 
{\bar \psi} ( \not \! \partial + i e \not \!\! A + M ) \psi \right]
\label{dfgt}
\ee
We are using Euclidean Dirac's matrices in the representation
$\gamma_{\mu} = \sigma_{\mu}$
and $\beta = \frac{1}{T}$ is the inverse temperature. The label
$3$ is used to denote the Euclidean time coordinate $\tau$.
The fermionic (gauge) fields in (\ref{dfgt}) obey  antiperiodic (periodic) 
boundary
conditions in the timelike direction.
We are concerned with the mass-dependent parity-odd piece 
$\Gamma_{odd}$ of $\Gamma$,
which, as a parity transformation changes the sign of the mass
term (the only odd term under parity in the Euclidean action), can be obtained
as follows:
\be
2 \Gamma_{odd} (A,M) = \Gamma(A,M)- \Gamma(A,-M)
\label{dfgo}
\ee
In any gauge invariant regularization scheme there is also a
mass-independent (and temperature independent) contribution (the parity 
anomaly) which corresponds to 
a CS term with a coefficient such that it changes in multiples of 
$i\pi$ under large gauge transformations \cite{dgs,GRS}.

The calculation of (\ref{dfgo}) for
{\em any \/} gauge field configuration is not something we can
do exactly. Instead of making a perturbative calculation dealing
with a small but otherwise arbitrary gauge field configuration,
we shall consider a restricted set of gauge field configurations
which can however be treated exactly. Moreover, as we want to make a
calculation which preserves the symmetry for gauge transformations with
non-trivial winding around the time coordinate, any approximation
assuming the smallness of $A_3$ could put this symmetry in jeopardy.

A convenient class of configurations from this
point of view is that of time-independent magnetic
fields  in a gauge such that
\be
A_3 = A_3 (\tau), \;\; A_j = A_j (x) \,\,\,  (j=1,2)
\label{rset}
\ee
namely, $A_3$ is only a function of $\tau$, and $A_j$ is
independent of $\tau$. Under these assumptions, we see that the only
$\tau$-dependence of the Dirac operator comes from $A_3$. This
dependence can however be erased by a redefinition of the integrated
fermionic fields.
The set of allowed gauge transformations in the
imaginary time formalism is defined in the usual way:
$$\psi (\tau,x) \; \to \; e^{-i e \Omega (\tau,x)} \psi (\tau,x) \;\;,\;\;
{\bar \psi} (\tau,x) \; \to \; e^{i e \Omega (\tau,x)} {\bar \psi}
(\tau,x)$$
\be
A_\mu (\tau,x) \;\to \; A_\mu (\tau,x) \,+\, \partial_\mu \Omega (\tau,x)
\ee
where $\Omega(\tau,x)$ is a differentiable function vanishing at spatial
infinity
($|x| \to \infty$), and whose time boundary conditions are chosen in order
not to affect the fields' boundary conditions.
It turns
out that $\Omega(\tau,x)$ can wind an arbitrary number of times around
the cyclic time dimension:
\be
\Omega(\beta,x) \;=\; \Omega(0,x) \,+\, \frac{2 \pi}{e} \, n
\label{bcomega}
\ee
where $n$ is an integer which labels the homotopy class of the gauge
transformation. 

As we are interested in evaluating the fermionic
determinant in a gauge invariant way
\be
\det ( \not \! \partial + i e \not \!\! A \,+\, M ) \; = \; \int {\cal D}
\psi \, {\cal D} {\bar \psi}
\; \exp \left\{ - \int_0^\beta d \tau \int d^2 x {\bar \psi} ( \not \!
\partial + i e \not \! A \,+\, M )
\psi \right \} \;,
\label{fdet}
\ee
we can always perform a gauge transformation 
in order to pass to an equivalent expression where the gauge field $A_\mu'
=
A_\mu + \partial_\mu \Omega$ is constant in time.
For the particular set of configurations (\ref{rset}) such a transformation
renders $A_3'$ constant.
We see that there is a family of $\Omega$'s achieving this while respecting
the boundary conditions (\ref{bcomega}),
\be
\Omega (\tau) \;=\;
- \int_0^\tau d {\tilde \tau} A_3 ({\tilde \tau}) +
\left( \frac{1}{\beta} \int_0^\beta d {\tilde \tau} A_3 ({\tilde \tau})
+\frac{2\pi n}{e\beta} \right) \tau
\ee
where $n$ is an arbitrary integer.
The freedom to chose $n$ could be used to further restrict the
values of the constant $A_3'$ to a finite interval. 
In this sense, the value of the constant
in such an interval is the only `essential', i.e., gauge invariant,
$A_3$-dependent information
contained in the configurations (\ref{rset}), describing the holonomy
$\int_0^\beta d {\tilde \tau} A_3 ({\tilde \tau})$. 
However, we will limit ourselves to small gauge
transformations ($n=0$) in order to avoid any assumption about large gauge
invariance of the fermionic measure in (\ref{fdet}) and safely discuss the
effect of large gauge transformations on the final results. Thus the constant
field $A_3'$ simply takes the mean value of $A_3(\tau)$, 
$ {\tilde A}_3 \;=\; \frac{1}{\beta} \, \int_0^\beta \, d \tau \, A_3 (\tau)$.
Note that the spatial components of $A_\mu$ remain $\tau$-independent
after this transformation.

It is convenient to perform a Fourier
transformation on the time variable for $\psi$ and ${\bar \psi}$, since the
Dirac operator is now invariant under translations in that coordinate:
\be
\psi (\tau, x) = \frac{1}{\beta} \, \sum_{n=-\infty}^{n=+\infty} \,
e^{i \omega_n \tau} \psi_n (x), \;\;
{\bar \psi} (\tau, x) = \frac{1}{\beta} \, \sum_{n=-\infty}^{n=+\infty} \,
e^{-i \omega_n \tau} {\bar \psi}_n (x) 
\ee
where $\omega_n = (2 n +1) \frac{\pi}{\beta}$ is the usual Matsubara
frequency for fermions.
Then the Euclidean action is written as an infinite series of decoupled
actions, one for each Matsubara mode
\be
 \frac{1}{\beta}
\sum_{n=-\infty}^{+\infty}
\int d^2 x {\bar \psi}_n (x) \left[ \not \! d \,+\, M \,+\, i \gamma_3
(\omega_n + e {\tilde A}_3) \right] \psi_n (x)
\ee
where $\not \!\! d=\;\gamma_j (\partial_j + i e A_j)$ is the Dirac operator 
corresponding to the spatial
coordinates and the spatial components of the gauge field.

As the action splits up into a series and the fermionic measure can be written
as
\be
{\cal D} \psi(\tau,x) \, {\cal D} {\bar \psi}(\tau,x)=
\prod_{n=-\infty}^{n=+\infty}{\cal D} \psi_n(x) \, {\cal D} {\bar \psi}_n(x)
\label{measure}
\ee
the $2+1$ determinant is an infinite product of the corresponding $1+1$
Euclidean Dirac operators
\be
\det ( \not \! \partial + i e \not \! A \,+\, M ) \; = \;
\prod_{n=-\infty}^{n=+\infty} \det [\not \! d + 
\rho_n e^{i \gamma_3 \phi_n} ] \; ,
\label{xx}
\ee
where we have also defined
\be
\rho_n \;=\; \sqrt{ M^2 + ( \omega_n + e {\tilde A}_3 )^2 }\;;
\phi_n \;=\; {\rm arctan} ( \frac{\omega_n + e {\tilde A}_3}{M} ) \;.
\ee
Explicitly, the $1+1$ determinant for a given mode is a functional integral
over $1+1$ fermions
\be
\det [\not \! d + \rho_n e^{i \gamma_3 \phi_n} ]  = 
 \int {\cal D} \chi_n \, {\cal D} {\bar \chi}_n \;
\exp\left\{ - \int d^2 x {\bar \chi}_n (x)  
( \not \! d + \rho_n e^{i \gamma_3 \phi_n} )  
\chi_n (x) \right\} \;.
\label{fori}
\ee
We now realize that the change of fermionic variables
\be
\chi_n (x) \;=\; e^{- i \frac{\phi_n}{2} \gamma_3} {\chi'}_n (x) \;\;,\;\;
{\bar \chi}_n (x) \;=\; {{\bar \chi}'}_n (x) e^{- i \frac{\phi_n}{2}
\gamma_3} \;,
\label{chiral}
\ee
makes the action in (\ref{fori}) independent of $\phi_n$. Concerning
the fermionic measure, it picks up an
anomalous Fujikawa jacobian \cite{fuji} so that one ends with
\be
\det [\not \! d + M + i \gamma_3 (\omega_n
+ e {\tilde A}_3) ] \;=\; J_n \; \det [\not \! d +  \rho_n ]
\label{fuji}
\ee
where
\be
J_n \;=\; \exp ( -i \frac{e \phi_n}{2 \pi} \int d^2 x \epsilon_{jk}
\partial_j A_k )\;,
\ee
with $\epsilon_{jk}$ denoting the $1+1$ Euclidean Levi-Civita symbol.

Recalling the definition of $\Gamma_{odd}$, we see that the second factor
in expression (\ref{fuji}) does not contribute to it, since it is
invariant under $M \to -M$. As a consequence, the parity odd
piece of the effective action is given in terms of the infinite set
of $n$-dependent Jacobians:
\be
\Gamma_{odd} \;=\; - \sum_{n=-\infty}^{n=+\infty} \, \log J_n
         \;=\; i \frac{e}{2 \pi} \, \sum_{n=-\infty}^{n=+\infty}
\phi_n \; \int d^2 x \epsilon_{jk} \partial_j A_k \;.
\ee
There only remains to perform the summation over the $\phi_n$'s,
whose sign will obviously depend on the sign of $M$.
Using standard finite-temperature techniques, this series
can be exactly evaluated, yielding
\be
\Gamma_{odd} \;=\; i \frac{e}{2\pi} 
{\rm arctan} \left[ {\rm tanh}(\frac{\beta M}{2}) 
\tan ( \frac{e}{2} \int_0^\beta d \tau A_3(\tau) ) \right]
\, \int d^2 x \epsilon_{jk} \partial_j A_k \;.
\label{espl'}
\ee
This is one of the main results in our paper: we
have been able to compute the {\it exact} mass-independent parity-odd 
piece of the effective action in $QED_3$  at finite temperature for the 
restricted set of configurations (\ref{rset}). Several
important features of this result should be stressed.

First of all, this result has the proper 
zero temperature limit:
\be  
\lim_{T \to 0} \Gamma_{odd} =\frac{i}{2}  \frac{M}{|M|}S_{CS}=
\frac{i}{2}  \frac{M}{|M|}
\frac{e^2}{2\pi}\int d^3x \epsilon_{\mu\nu\alpha}A_\mu
\partial_\nu A_\alpha .
\label{T=0}
\ee
As it is well known, in the zero temperature case the mass-dependent
part of the parity breaking is not
invariant under large gauge transformations. The quantization of
the flux of the magnetic field in the last factor of (\ref{espl'})
as  $q\frac{2\pi}{e}$ shows that (\ref{T=0}) changes 
by the addition of $i nq\pi$
under a large gauge transformation (\ref{bcomega}). The gauge 
non-invariance appearing when $n$ and $q$ are odd is compensated by
the parity anomaly when the result is
regularized in a gauge invariant scheme.
Notice also that only in the zero temperature limit is the result 
non-analytic in $M$.

The same situation occurs in the finite temperature result (\ref{espl'}).
A large gauge transformation with odd winding number $n=2p+1$ shifts the 
argument of the tangent in $(2p+1)\pi$. Although the tangent is not 
sensitive to such a change, one has to keep track of it by shifting the 
branch used for arctan definition. This amounts to exactly the same result 
as in the $T \to 0$ limit.

Next we observe that an expansion in powers of $e$ yields the usual
perturbative result
\be
\Gamma_{odd} = \frac{i}{2}  \tanh(\frac{\beta M}{2})S_{CS}
+O(e^4)
\label{pert}
\ee
where the coefficient of the Chern-Simons term acquires a smooth temperature
dependence. Were we considering only the first non trivial order in $e$,
we would find a clash between temperature dependence and gauge invariance
\cite{bfs}-\cite{cfrs}. Now we learn, as it was first stressed in \cite{dll}
in a $0+1-$dimensional example and in \cite{dgs} in a setting similar to ours,
that one has to consider the full result in order to analyze gauge invariance.
The apparent impossibility of respecting gauge invariance shown by 
(\ref{pert}) is in fact compensated by non-local higher order terms in 
the perturbative expansion. 

Finally, we observe that the result (\ref{espl'}) is not an extensive quantity
in Euclidean time. It is however extensive in space, and that is indeed
all one expects in Finite Temperature Field Theory. In contrast, the $T=0$
limit becomes an extensive quantity in space-time, as is expected from zero
temperature Field Theory. 

We shall now extend the previous results to the somewhat more general
situation of gauge fields satisfying the constraint of $A_j$ being
again time-independent, but allowing for a smooth spatial dependence
of $A_3$  besides the previous arbitrary time dependence.

The fermionic determinant we should calculate, after getting rid of the
$\tau$ dependence of $A_3$ will have a form analogous to (\ref{xx})
with the only difference of having an $x$ dependence in 
$\rho_n$ and $\phi_n$.
The determinant corresponding to the $n$-mode is again written as
in eq.(\ref{fori}) and we can perform the two-dimensional chiral rotation
(\ref{chiral}). The $x$-dependence of the phase factor $\phi_n$
produces in this case a different anomalous Jacobian,
\be
\det \left[ \not \!\! d + \rho_n (x)
e^{i \gamma_3 \phi_n (x)} \right] \;=\; J_n' \; \det [\not \! d' +  \rho_n (x)] \;,
\label{fact}
\ee
where $\not \! d' \;=\; \not \! d \,-\, \frac{i}{2} \not \! \partial \phi_n
\gamma_3 \;.
$
This affects the result in two
ways: first, as the fermionic operator in the r.h.s.\ 
depends on the sign of $M$, 
there will be a contribution to $\Gamma_{odd}$ coming from the determinant
of $\not \! d' +  \rho_n (x)$. Second, the Jacobian is a slightly 
more involved function of $\phi_n$\cite{fuji},
\be
J_n' \;=\; \exp \left\{-i \frac{e }{2 \pi} \int d^2 x [ \phi_n (x)
\epsilon_{jk}
\partial_j A_k + \frac{1}{4} \phi_n (x) \Delta \phi_n (x) ] \right\} \;.
\ee
In a first approximation,
we shall only take into account the contribution coming from the Jacobian,
since the one that follows from the determinant of the Dirac operator is of
higher order in a derivative expansion (and we are assuming that the
$x$-dependence
of ${\tilde A}_3$ is smooth).
Moreover, the contribution which is quadratic in $\phi_n$ 
is irrelevant to the
parity breaking piece, since it is invariant under the change $M \to -M$.
Thus, neglecting the terms containing derivatives of ${\tilde A}_3$, we
have for $\Gamma_{odd}$ a natural generalization
of eq.(\ref{espl'})
\be
\Gamma_{odd} \;=\; i \frac{e}{2 \pi} \int d^2 x
\; {\rm arctan} \left[ \tanh(\frac{\beta M}{2}) 
\tan ( \frac{e}{2} \int_0^\beta d \tau
A_3(\tau, x) ) \right]\,  \epsilon_{jk} \partial_j A_k (x)\;.
\ee
The approximation of neglecting derivatives of ${\tilde A_3}$ is 
reliable if the condition
$| e \; \partial_j {\tilde A_3} | \;<<\; M^2$
is fulfilled.
To end with this example, let us point that all
the remarks we made for the case of a space-independent $A_3$ 
apply also to this case.

In conclusion, using  particular gauge field configurations we have
computed the mass-dependent parity-violating part of the
effective action for $2+1$ massive fermions at finite
temperature obtaining an exact result. 
Once the standard parity anomaly is taken into account, we have shown 
that gauge
invariance holds even when large gauge field configurations are
considered. Our method, which made use of the calculation of 
the $1+1$ anomaly, can be also applied to the analysis of the
non-Abelian case; details of this case will be given elsewhere.

\underline{Acknowledgements}: The authors would like to thank S.Deser,
L.Griguolo and D.Seminara for discussions and corrections on the original 
manuscript. G.L.R. thanks R.Jackiw for kind 
hospitality at CTP, MIT. C.D.F. and G.L.R. are supported by CONICET,
Argentina.
F.A.S. is
partially  suported
by CICBA, Argentina and a Commission of the European Communities
contract No:C11*-CT93-0315. This work is 
supported in part by funds provided by the U.S. Department of Energy (D.O.E.)
under cooperative research agreement \# DF-FC02-94ER40818.

\end{document}